\documentclass{article}
\usepackage{graphicx} 
\usepackage[table]{xcolor}
\usepackage{color, colortbl}

\usepackage{authblk}
\providecommand{\keywords}[1]{\textbf{\textit{Keywords:}} #1}

\title{The Role and Applications of Airport Digital Twin in Cyberattack Protection during the Generative AI Era} 

\author[1]{Abraham Itzhak Weinberg}
\affil[1]{AI-WEINBERG, AI Experts, Tel Aviv, Israel, aviw2010@gmail.com}

\begin{document}
\maketitle
\begin{abstract}
In recent years, the threat facing airports from growing and increasingly sophisticated cyberattacks has become evident. Airports are considered a strategic national asset, so protecting them from attacks, specifically cyberattacks, is a crucial mission. One way to increase airports' security is by using Digital Twins (DTs). This paper shows and demonstrates how DTs can enhance the security mission. The integration of DTs with Generative AI (GenAI) algorithms can lead to synergy and new frontiers in fighting cyberattacks. The paper exemplifies ways to model cyberattack scenarios using simulations and generate synthetic data for testing defenses. It also discusses how DTs can be used as a crucial tool for vulnerability assessment by identifying weaknesses, prioritizing, and accelerating remediations in case of cyberattacks. Moreover, the paper demonstrates approaches for anomaly detection and threat hunting using Machine Learning (ML) and GenAI algorithms.
Additionally, the paper provides impact prediction and recovery coordination methods that can be used by DT operators and stakeholders. It also introduces ways to harness the human factor by integrating training and simulation algorithms with Explainable AI (XAI) into the DT platforms. Lastly, the paper offers future applications and technologies that can be utilized in DT environments.
\end{abstract}


\section{Introduction}
Airports have increasingly become targets of sophisticated cyberattacks in recent years, facing growing threats to their critical digital infrastructure and passenger safety systems from both nation-state actors and criminal hacker groups \cite{lykou2018smart}.
A Digital Twin (DT) of an airport's full operations can play an important role in defending against cyberattacks by providing a secure virtual environment to model threats, test mitigations, and help ensure continuity of critical services \cite{kuleshov2024cyber,grieves2017digital}.
Generative Artificial Intelligence (GenAI) has emerged as a disruptive approach in recent years, garnering increasing recognition. It has the potential to serve as a powerful virtual tool within the airport DT, enabling the modeling and testing of potential cyberattacks in sophisticated simulated scenarios \cite{xu2024leveraging}.
This paper introduces the potential of DT as an assisting tool for airport applications pertaining to cyberattacks in general. It also highlights the advantages of utilizing GenAI within the DT context.\\ 
DT is defined as a virtual representation of a physical product, system, or process in the real world \cite{vanderhorn2021digital}. It serves as an accurate and interchangeable digital counterpart for various practical purposes, including simulation, integration, testing, monitoring, and maintenance \cite{lyu2024handbook}. A DT for Airport Operations Centre (APOC) in airports is a virtual representation of the entire airport ecosystem, encompassing its infrastructure, facilities, systems, and operations.
GenAI refers to the capability of Artificial Intelligence (AI) to produce text, images, videos, or other data through the use of generative models \cite{feuerriegel2024generative}. These models learn the patterns and structure of the input training data and subsequently generate new data that exhibits similar characteristics. GenAI is often prompted to create content based on specific inputs, allowing it to generate unique and relevant outputs \cite{airportir}.

\subsection{Background on Cyber Threats to Airports}
Airports face growing and increasingly sophisticated threats due to their interdependence, digitization and role as nerve centers of regional connectivity.
They are being attractive targets due to the vast amounts of sensitive data processed and the need to ensure smooth operations for safety, face various threats from sophisticated state actors to opportunistic criminal groups \cite{lykou2018smart}. The pandemic lockdowns further exacerbated vulnerabilities as unpatched systems increased and remote work amplified phishing risks, often leading to unreported attacks \cite{elmarady2022impact}. The interconnectedness of Information technology (IT) / Operational Technology (OT) systems in airports means that a single breach can impact multiple tenants and partners, including airlines and transportation systems. However, the lack of standardized security practices across different airport authorities makes defending the ecosystem challenging and leaves individual gaps vulnerable. \\ Geopolitical conflicts and tensions may also involve targeting airport infrastructure to cause economic or reputational damage to a nation. Despite airports' rising importance as logistical hubs, cyberdefense often takes a backseat to operational priorities, leaving them exposed. Legacy systems pose a unique challenge as they are difficult to patch and control, yet remain critical. The main cyber threats that airports face include ransomware attacks, destructive malware, Denial-Of-Service (DOS) and  Distributed Denial-Of-Service (DDoS) attacks, phishing and social engineering, insider threats, supply chain compromise, data exfiltration, ramp equipment hijacking, cyber-physical attacks, and disinformation campaigns \cite{koroniotis2020holistic,auer2019cyber}. It is crucial to address these threats to ensure the resilience and smooth functioning of airports as national security priorities.\\
In the past decade especially, many of the world's largest airports have faced disruptive or data-stealing cyberattacks largely due to their lack of coordinated digital defenses \cite{senarak2024port,zhiling2024construction}.
There have been several significant cyberattacks on airports, highlighting the vulnerabilities they face in today's digital landscape. In October 2022, the online operations of over a dozen U.S. airport websites were compromised due to cyberattacks, which have been attributed to the Russian hacker group Killnet. In 2021, Amsterdam's Schiphol Airport experienced a ransomware attack that resulted in flight delays and check-in problems, with the attackers threatening to leak private data. In 2020, Los Angeles International Airport fell victim to cybercriminals who hijacked the public address system, broadcasting false evacuation orders to induce panic. London's Gatwick and Birmingham airports faced disruptions in 2019 when a malware infection affected digital signage and other displays, causing flight information issues for passengers.\\ The world's busiest air travel period in 2018 was marred by a ransomware attack on Atlanta's Hartsfield-Jackson International Airport, leading to disruptions in check-in systems during the Thanksgiving holiday. In 2017, San Diego International Airport experienced a data breach that compromised the personal and financial information of 3.5 million individuals, attributed to an email phishing campaign. In 2016, security vulnerabilities were uncovered in the networks of New York's John F. Kennedy Airport and other major US airports, raising concerns about potential access to critical systems. In 2015, cyberespionage hackers targeted airline carrier networks, stealing data on 25 million passengers from five different carriers over an 18-month period. Finally, in 2013, Los Angeles International Airport fell victim to a hack that planted malware, linked to an advanced threat group with a focus on targeting US infrastructure. These incidents serve as stark reminders of the pressing need for robust cybersecurity measures in the aviation industry.

\subsection{Overview of Airport Digital Twins and Generative AI Capabilities}
Airport DTs provide a detailed and comprehensive overview of airport operations by collecting and integrating real-time data from various OT/IT systems such as security cameras, weather sensors, baggage handling equipment, and check-in kiosks. This data is used to create a virtual replica of the entire airport facility, mirroring the physical airport in the digital world \cite{nath2021building,digitalplatform}. The DT remains synchronized with real-time changes, ensuring its accuracy as a dynamic model.\\ Airport operators can visualize, analyze, and optimize complex operations, such as passenger flow modeling and resource allocation, using the DT. The simulation capabilities of DT enable the testing of "what if" scenarios for planned projects, upgrades, and emergency responses without disrupting live operations. The insights gained from the twin support data-driven decision making, aiding in CAPital EXpenditures (CAPEX) \footnote{Costs accrued for obtaining tangible assets intended for long-term utilization.}/OPErating EXpenses (OPEX) \footnote{Costs associated with sustaining the daily functioning of a company.} planning, schedule coordination, and more \cite{mcclenaghan2023cost}.\\
Advanced DTs leverage technologies like AI and Machine Learning (ML) and Augmented / Virtual Reality (AR/VR), to understand the relationships between different systems and key performance indicators \cite{yin2023state, lv2021artificial, uzun2019design}. Performance can be virtually benchmarked and optimized across metrics like throughput, energy usage, and maintenance. The integration of DTs helps airports transition toward more connected and sustainable models for the future, leading to accelerated innovation and improved passenger experiences. In this context, GenAI capabilities take airport DTs to the next level. Through integration with GenAI, DTs acquire the capability to simulate unforeseen and innovative scenarios that go beyond typical "what if" analyses, as mentioned earlier. The combination of DT modelling with GenAI's out-of-the-box thinking enables stakeholders to explore creative scenarios and gain new insights that were previously unimagined. This characteristic enables the airport to be better prepared, not only for past attack scenarios and vectors, but also for unforeseen ones.\\
GenAI automatically simulates complex scenarios based on statistical modeling of real behaviors and events. Techniques like Generative Adversarial Networks (GANs) and Transformers can synthetically generate virtual passengers, vehicles, and even cyberattacks within the virtual airport environment. This expands the capabilities of DTs, providing a large and flexible testbed for researching defenses against AI and adaptive threats. Simulations can be customized based on changes to conditions, policies, or external threats using conditional process modeling and machine translation. DTs enhanced with GenAI function as living laboratories, compressing years of real-world activity and risks into compressed timeframes. The insights gained from analyzing the outcomes of numerous simulations strengthen decision making regarding security investments and incident response planning for airports \cite{saifutdinov2020digital}. DTs serve as risk-free virtual spaces where airports can test innovations and manage changes and disruptions, while GenAI enhances their capabilities and potential.

\section{Modeling Cyberattack Scenarios}
An airport DT possesses a crucial capability of modeling realistic cyberattack scenarios, enabling security teams to simulate potential threats within a closed and virtual environment prior to their occurrence in the real world \cite{cheng2023review}. This approach allows for the identification of vulnerabilities and evaluation of defenses that would be challenging or risky to test against live production systems. The DT can emulate various common scenarios, including targeted intrusions, ransomware and malware outbreaks, insider threats, supply chain compromises, and the physical consequences resulting from cyber-physical attacks \cite{auer2019cyber}. It facilitates the simulation of emerging threats that involve a combination of cyber and physical elements, such as the hijacking of airport systems or coordinated multi-vector attacks. By implementing intelligent adversarial behavior programming, the DT can autonomously execute sophisticated attacks that closely resemble real-world tactics, demonstrating advanced strategic maneuvers \cite{ding2024generative}. Observing the outcomes of these simulated attacks provides valuable insights, such as identifying cascading impacts or unanticipated consequences, which in turn inform the prioritization of remediation efforts. Additionally, the DT serves as a dynamic laboratory for prototyping new defenses before their implementation, enabling the measurement of their effectiveness against evolving cyber criminality. 

\subsection{Simulating Common and Emergent Attack Vectors}      As mentioned above, an airport DT can simulate various common and emerging cyberattack vectors, providing valuable insights and enhancing security measures. It can simulate ransomware, phishing, and malware attacks on different airport systems, such as check-in, security, and baggage handling, to analyze potential impacts and vulnerabilities \cite{souanef2023digital}. Additionally, it can model DoS or Distributed DDoS attacks targeting public-facing websites, WiFi networks, or operational systems, assessing their potential consequences. The DT can also test "insider threat" scenarios, including compromised credentials or malware spreading through contractors \cite{su2023using,kuleshov2024cyber}. Simulating emerging vectors like hacking Internet of Things (IoT) or OT devices and integrating systems like digital displays and security cameras helps evaluate potential risks \cite{koroniotis2021sair}. Defenses can be stress-tested against sophisticated targeted attacks, including zero-day exploits and supply chain compromises. By generating synthetic phishing emails, ransomware variants, or destructive malware payloads, previously undiscovered vulnerabilities can be identified. The DT can dynamically change the Tactics, Techniques, and Procedures (TTPs) of digital adversaries to simulate real-world cybercriminal behaviors \cite{papadopoulos2024protection}. Leveraging AI, it can autonomously pilot automated cyberattacks that learn, adapt, and escalate their behaviors over multiple simulations. A/B testing \footnote{A/B testing, also called split testing, is a method to compare two versions of a webpage or app and determine their relative performance.} mitigation strategies under various situational conditions helps identify gaps and optimize security postures. Ultimately, the airport DT provides a safe and flexible environment to prototype and experiment with combating future forms of cybercrime.

\subsection{Generating Synthetic Threat Data for Testing Defenses}
An airport DT has the capability to generate synthetic threat data, which can be utilized for testing cyber defenses in a controlled environment \cite{xiong2022digital}. By employing generative models like GANs or Diffusion models, the DT can create realistic fake credentials, documents, files, and traffic that appear authentic but are actually simulated attacks \cite{saifutdinov2020digital}. It can synthesize new variants of malware, phishing emails, and ransomware payloads based on analyzed patterns from recent incidents, incorporating novel coded behaviors. Additionally, the DT can generate synthetic network traffic that replicates the tactics, techniques, and procedures of known threat actors, integrating internally-created adversarial routines. ML can be leveraged to profile past incidents and generate simulated timelines of multi-stage attacks involving combinations of vectors that have not been seen before. Training generative language models enables the crafting of disinformation campaigns on social media, capable of sowing fear or disrupting travel if not promptly addressed \cite{stoddart2022cyberwar}. The DT can also generate digital identities, social media accounts, and online histories for virtual individuals and devices, serving as entry points in simulated attacks. To establish realistic baseline usage and traffic patterns, the DT can synthesize sensor logs, system configurations, and other airport IT/OT data. Furthermore, it can continuously expand its simulated threat repertoire using generative models, ensuring that defenses are consistently challenged by emerging tools and techniques. This approach enables the creation of an endless stream of contextualized test scenarios, facilitating the evaluation of control improvements within the airport DT.

\subsection{Scenario Testing with Realistic Impacts on Operations}
An airport DT offers several ways to test cyberattack scenarios and assess their realistic impacts on airport operations, enabling the prioritization and validation of mission-critical system defenses \cite{yiu2021digital,conde2022applying}. Through simulation, the DT can replicate a ransomware attack targeting check-in systems during a busy travel period, allowing for the observation of flow-on delays, diversions, and impacts on passengers \cite{brucherseifer2021digital}. As mentioned before, it can also model a dedicated DDoS attack that overwhelms the airport's WiFi network, providing insights into reduced productivity and disruptions in safety communications. Testing a potential breach in the air traffic control system allows for the evaluation of risks related to flight plan manipulation or navigation disruptions. Additionally, the DT can stage an attack on the baggage handling infrastructure to analyze the consequential flight cancellations resulting from delayed luggage transportation. By running scenarios involving the breach of security camera feeds or access control systems, the potential safety and criminal impacts can be measured. The DT can also test emergent attacks that combine cyber and physical risks, such as compromised aircraft fueling systems or ground vehicle hijacking.\\ Continuous simulation of "day-in-the-life" airport operations under realistic cyber stress testing contributes to the refinement of resilience strategies. Quantitative comparisons can be made between disruption impacts and recovery timelines under different attack vectors, such as rapidly spreading malware versus targeted intrusions. Furthermore, the DT can track the dynamics of cascading or secondary effects across interdependent airport networks and partner organizations. Ultimately, these capabilities provided by the airport DT aid in prioritizing and validating the defense of the most critical systems.

\section{Vulnerability Assessment and Patch Development}
An airport DT offers valuable capabilities for vulnerability assessment and patch development, enabling the identification of weaknesses and prioritization of remediation efforts through virtual penetration testing and predictive risk analysis. With its granular modeling of the airport infrastructure and systems, the DT serves as a testbed for red team exercises, allowing for thorough evaluations of digitized assets and the discovery of vulnerabilities, misconfigurations, and gaps in existing security controls that could be exploited \cite{piroumian2023cybersecurity}. Critical issues that pose a threat to safety or business continuity are promptly flagged for immediate attention, while others are risk-scored based on their likelihood and potential impact. Developers can leverage the DT to rapidly prototype, test, and validate patches in an isolated replica of the production environment, ensuring that fixes do not inadvertently disrupt authorized users or introduce new flaws before being rolled out. By continuously assessing evolving risks, the DT also guides strategic roadmaps and technology selection, enabling the fulfillment of long-term protection needs. Through its simulation capabilities, the DT accelerates remediation timelines and ensures that updates effectively address future threats without causing disruptions. By integrating vulnerability management into operational planning, airports can minimize their exposure to risks by implementing predictive and proactive security measures.

\subsection{Identifying Weaknesses through Virtual Penetration Trials}
An airport DT offers various methods for identifying vulnerabilities through virtual penetration testing \cite{li2021digital}. It can model common internet-facing systems and conduct simulated external ethical hacking attempts using tools such as Network Mapper (Nmap), Nikto, and Metasploit \cite{faleiro2022digital,wen2022toward}. Additionally, the DT can clone internal networks and perform internal penetration tests to emulate advanced insider threats or supply chain attacks \cite{kaewunruen2021digital}. It is also capable of stress-testing \footnote{Stress testing involves rigorously testing a system, infrastructure or entity beyond its normal operational capacity to assess its stability under extremely challenging conditions \cite{padovano2024improving}. The goal is to observe how it performs and reacts when pushed to or beyond its limits, often to the point of failure. This provides insight into weaknesses or vulnerabilities that may not be apparent during standard testing parameters.} the authentication, authorization, and session management of applications to uncover vulnerabilities like credential stuffing. Testing Application Programming Interfaces (APIs), Software Development Kits (SDKs), and adjacent partner/vendor systems allows for the detection of injection flaws like SQL injection (SQLi) and Cross-Site Scripting (XSS) through simulated bot-driven exploits \cite{borras2020d1}. Furthermore, the DT can leverage AI to autonomously discover and chain together zero-days and unknown vulnerabilities that have not been observed before. By mining unstructured data sources such as documents, manuals, and schemas, the DT can identify unintended exposure of sensitive assets. Fuzz-testing \footnote{Fuzz testing, also known as fuzzing, is an automated software testing technique that injects abnormal inputs into a system to expose software flaws and weaknesses. By injecting these inputs and monitoring for exceptions like crashes or data leaks, a fuzzing tool uncovers potential vulnerabilities.} interfaces, protocols, and legacy/IoT devices with chaotic inputs aids in the identification of memory corruption bugs \cite{gao2022adequate}. Accurately modeling digital asset configurations enables the identification of misconfigurations that could serve as vectors of attack. The DT can also replay historic breaches against hypothetical modern infrastructure to assess residual risks. Lastly, by continuously syncing tests with real-world TTPs, the DT validates dynamic security measures against current and future threats.

\subsection{Prioritizing Remediation Through Predictive Risk Analysis}
An airport DT plays a crucial role in prioritizing vulnerability remediation through predictive risk analysis. By utilizing attack graphing techniques, the DT can map the potential pathways an adversary could take from vulnerabilities to high-value assets, providing valuable insights for prioritization \cite{sadeghi2024digital}.\\ Probabilistic modeling is applied to estimate the likelihood of exploitation based on historical indicators such as Common Vulnerability Scoring System (CVSS) scores. Through simulation of different attack scenarios, the DT analyzes the potential business impacts of vulnerability compromise, taking into account criticality to safety, scheduling, and revenue, in order to determine severity and risk ratings. Continuously rescoring vulnerabilities based on emerging threats ensures that patching efforts are prioritized to address emerging risks. The DT employs advanced modeling of multi-vector attacks to predict the viability of systems if breached, enabling proactive measures \cite{nsoh2021next}. The effectiveness of existing controls can be assessed through simulation, even after mitigation, by retaining vulnerabilities and evaluating their impact. Recommendations for technical and policy remedies, as well as advanced monitoring of high-risk assets, are provided by the DT. Additionally, the DT analyzes the costs, benefits, and risks associated with patching, including temporary workarounds versus permanent fixes, to guide investment decisions. Vulnerabilities impacting systems integral to incident response capabilities are given high priority, ensuring proactive remediation of the most mission-critical weaknesses before they can be exploited.

\subsection{Accelerated Development and Validation of Fixes}
An airport DT offers several ways to expedite the development and validation of vulnerability fixes, ensuring efficient remediation timelines while minimizing risks and maintaining stable operations \cite{hazbon2019digital}. The DT can clone affected systems and networks, providing an isolated environment for developers to quickly prototype patches. Leveraging simulation capabilities, it enables the replay of exploitation attempts against patched variations, validating the effectiveness of the fixes. As mentioned above, A/B testing proposed solutions allow for the analysis of second-order impacts, such as performance or dependent process considerations. By modeling patches at scale across the entire digitally replicated infrastructure, the DT can identify conflicts or regressions early on. Regression testing \footnote{Regression testing re-runs tests on modified software to ensure it still functions as expected.} is automated by re-running comprehensive penetration tests with each iteration, ensuring that vulnerabilities remain closed \cite{kilic2023digital}. Simulating day-to-day usage and fine-tuning patches within the DT helps avoid any potential disruptions or introduction of new bugs. In case of problematic fixes, the DT allows for a rollback to prior validated versions before committing to production. Assessing long-term security improvements is facilitated through vulnerability retrospectives conducted over time with applied patches. Finally, by continuously synchronizing the DT with the codebase \footnote{a collection of computer code that comprises a particular software application, system or component. It includes all the source code files needed to build and release that specific software product or module.}, the readiness of all fixes for production deployment can be certified \cite{reitenbach2020collaborative}. In this way, the airport DT accelerates the development and validation of vulnerability fixes, reducing timelines, mitigating risks, and maintaining stable operations. 

\section{Training and Simulation Exercises}
An airport DT serves as a effective tool for conducting training and simulation exercises, allowing staff to prepare for potential incidents and emergencies in a highly realistic manner \cite{ruiz2024application}. By dynamically recreating conditions within the virtual replica, various scenarios can be staged and thoroughly studied in a risk-free environment. Trainees actively engage with the digitally simulated airport, responding to unfolding events, and their decisions and performance metrics are evaluated for analysis and improvement. These virtual simulations are instrumental in identifying gaps, validating emergency plans, and building crucial "muscle memory" before facing high-stakes situations in the real world. The DT enables the realistic modeling of different scenarios involving threats such as fires, extreme weather, aircraft crashes, and security breaches. Moreover, it facilitates the rehearsal of rare and unexpected "black swan" events, enabling teams to practice multi-agency coordination and on-site management. Debriefing sessions following the simulations provide valuable education on best practices, highlight areas for process improvement, and extract lessons learned. Overall, the airport DT creates a cost-effective and dynamic testbed that allows for continuous honing of staff skills in addressing complex scenarios, ensuring preparedness and readiness in the face of real-world challenges.

\subsection{Realistic Staff Training Environments}
An airport DT offers realistic staff training environments by incorporating various features and capabilities \cite{wong2023smart}. It digitally recreates the layout, infrastructure, and workflows of the airport to provide trainees with a familiar setting for immersion. The twin can also simulate ambient conditions such as weather, lighting, and crowds, establishing situational context for training scenarios. To enhance realism, the DT models the behaviors and interactions of non-player characters like passengers, crews, and traffic. By integrating replicas of actual airport systems, sensors, and equipment, trainees can engage in hands-on practice with responding to different situations. The DT triggers dynamic events based on trainee actions, creating an unfolding storyline that requires effective management \cite{hristov2023challenges}. It populates the twin with vast historical and real-time data, enabling context-aware decision making. AR/VR technologies can be leveraged to provide immersive first-person views, placing trainees in the heart of the action \cite{tran2022applications,yin2023state,wong2021closed}.\\ Staff access and actions are authenticated through replicated credentials and procedures, ensuring a realistic training experience. The DT continuously evolves situations based on collective and individual trainee performance, allowing for adaptive training scenarios. Simulations can be archived for detailed debriefs, assessment of skills transference, and refresher training purposes. Another aspect of DT that will be discussed later in the XAI section is the integration between the human factor and the simulated environment. DT not only enables the integration between humans and machines in cyberattack scenarios but also provides explanations and suggestions from various digital sources. Through these capabilities, the airport DT helps internalize roles, strengthen coordination muscle memory, and keep staff readiness skills razor sharp.

\subsection{Tabletop Exercises for Incident Response Planning}
An airport DT can facilitate tabletop exercises \footnote{A tabletop exercise is a gathering of essential personnel responsible for emergency management tasks. They come together in a safe and simulated environment to discuss different emergency scenarios.} for incident response planning, offering valuable benefits in strengthening coordination and refining plans well before real crises occur \cite{ojo2024innovative}. By populating the twin with data-driven incident timelines, After-Action Reports (AAR), and response templates, staff have a comprehensive reference for their exercises \cite{applebystrategic}. The DT allows for manual injection and real-time advancement of scenarios, enabling the testing of action plans and new procedures. It automatically generates cascading secondary issues and complications based on inputs, creating a realistic level of complexity to challenge the response teams. During the exercises, operational temperatures, coordination breakdowns, and resource gaps can be measured and discussed, revealing areas for improvement. Interactions and decision points can be recorded, facilitating the identification of planning, communication, or capability shortfalls. Debrief sessions can involve replaying vignettes while freeze-framing to highlight lessons learned and identify gaps in the response. The scenarios can be customized to address unique risks such as infrastructure faults, extreme weather events, or active threats. The DT can also incorporate outside experts to validate plans or challenge staff assumptions from different perspectives, bringing valuable insights to the process. Collaborative what-if planning sessions allow for brainstorming responses under worst-case scenarios or emerging threats \cite{luo4806351high}. Tabletop exercise sessions can be archived for ongoing process enhancement and to onboard new staff, ensuring a continuous improvement cycle. Moreover, the Digital Twin (DT) can assist in enhancing the staff's capabilities to comprehend and make real-time decisions based on multiple concurrent scenarios. This can be achieved through integration technologies leveraging big data and ML systems.
Through these capabilities, the airport DT strengthens coordination and helps refine incident response plans, enabling effective preparedness well in advance of real crises.

\subsection{Simulating Continuity of Operations After Attacks}
An airport DT can serve as a key tool for simulating the continuity of operations after cyberattacks or other disruptive incidents, enabling the establishment of resilient and data-validated plans to swiftly recover operations \cite{piroumian2023cybersecurity,souanef2023digital}. The DT can model the cascading impacts of attacks on networked airport systems and their dependencies, allowing for a comprehensive understanding of the evolving situation over time. By dynamically degrading the functionality of impacted assets such as check-in, baggage handling, or air traffic control, the twin can simulate the effects of the attack on critical components. Strains on communications, power, and staffing can be simulated if primary or backup resources are compromised, providing insights into potential vulnerabilities \cite{brucherseifer2021digital}.\\ Redundancy plans can be tested by triggering secondary support activations and observing any strains or delays in the recovery process \cite{meyer2020development}. The DT enables the evaluation of mobile, pop-up, or manual workarounds for maintaining critical services in the face of technological outages. Surge response capabilities, including contact center hold times, alternative triage sites, or rebooking queues, can be assessed to ensure effective response strategies \cite{agapaki2022airport}. Disaster recovery and business continuity strategies can be stress-tested under evolving disruption scenarios, validating their effectiveness \cite{ariyachandra2023digital}. A/B testing contingency options and observing their outcomes within the DT helps validate optimal recovery pathways \cite{casey2024real}. The twin can generate after-action reports and data-driven timelines, providing valuable insights for continually refining response improvements. Additionally, the DT enables training of multi-stakeholder coordination across agencies, airlines, and partners, fostering a unified approach to recovery efforts. Through these capabilities, the airport DT empowers airports to establish resilient plans and respond rapidly following cyberattacks or other disruptive incidents.

\section{Anomaly Detection and Threat Hunting}
An airport DT serves as a comprehensive and high-fidelity virtual representation of airport systems, assets, data, and operations \cite{sindiramutty2023autonomous}. Within this digital replica, continuous monitoring of behaviors across the environment takes place, allowing for enhanced situational awareness and proactive identification of active compromises \cite{sadeghi2024digital}. Analytical models are trained using the vast historical and real-time data flows within the twin, establishing a dynamic baseline of normal activities. Anomaly detection algorithms then identify emerging deviations from this baseline, serving as indicators of compromised assets or insider threats. Unusual data access patterns, atypical credentials, or processes are flagged as potential risks. Threat hunters leverage the DT to iteratively search for subtle hints of intrusions that may have evaded other defenses. By probing digitized networks, data, and systems, they piece together unexpected findings and uncover hidden risks. The insights gained from hunting activities not only inform security improvements but also aid in incident response. Over time, this iterative process shapes a learning tactic and improves the detection of stealthier threats based on enriched patterns of malicious behavior. The airport DT acts as a living information hub, facilitating a proactive approach to catching active compromises within airport operations and enhancing overall security.

\subsection{Leveraging AI/ML for Networked Threat Monitoring}
An airport DT's AI/ML capabilities offer numerous ways to enhance anomaly detection and threat hunting across networked environments \cite{wang2023digital}. ML models can be deployed to analyze logs and flows throughout the digitized infrastructure, establishing baselines of expected communications \cite{hakiri2024comprehensive}. By continuously monitoring live data streams, the DT can flag statistically significant deviations from these baselines, indicating potential internal or external threats. Anomalies can be correlated across systems, locations, and timestamps to identify compromised or malicious internal actors moving laterally through the network. The DT can also recognize similarities between new deviations and historic incidents, enabling the rapid classification and prioritization of unknown risks. Clustering unexplained anomalies generates investigative hypotheses for human analysts to further probe via the twin. As ML algorithms continuously learn over time, they improve their accuracy in differentiating malignant behavior from false positives. AI assists with trace evidence analysis by automatically piecing together fragmented indicator data and artifacts, aiding in the identification of potential threats \cite{nath2021building}. Predictive analytics can anticipate how threats may evolve and provide insights on where to focus monitoring efforts, potentially intercepting evolving attacks earlier. The airport DT's AI capabilities supplement and scale 24/7 manned monitoring of the dynamic airport attack surface, providing critical analysis to enhance security measures.

\subsection {Proactive Identification of Intrusion Behaviors}
An airport DT's anomaly detection and threat detection capabilities offer various ways to proactively identify intrusion behaviors within the airport environment \cite{kuleshov2024cyber}. By establishing baselines of normal user/entity behaviors and developing profiles, the twin can detect abnormal account activity and flag anomalous login times or locations. ML can be utilized to analyze network traffic patterns, enabling the identification of tunneling, lateral movement attempts, or encrypted communications from blacklisted endpoints. The DT can inspect files, APIs, and data accessed for unusual data spills, privilege escalations, or modification of sensitive resources. Continuous authentication of asset configurations allows for the detection of suspicious changes that may indicate tampering with systems, applications, or network devices \cite{khan2022digital}. Monitoring filesystem events, disk usage, and memory consumption helps identify signs of hidden processes, shelled accounts, or dropped tools. The twin can detect irregular service restarts, stopped logs, and process injections that could potentially mask intruder activities. By setting up honeypots, monitoring dark web chatter, and correlating indicators, the DT aids in actively detecting domestic and foreign threats targeting the airport. It can also spot odd queries, access patterns, or anomalous data downloads to airgapped or restricted databases and systems \cite{susila2020impact}. Additionally, abnormal traffic from authorized planes or equipment can be monitored to identify potential supply chain threats, which can then be further analyzed. These capabilities of the airport DT significantly enhance the airport's ability to proactively identify and respond to intrusion behaviors, bolstering overall security.

\subsection {Advanced Analytics for Unknown Threat Profiling}
Advanced analytics in an airport DT offer valuable capabilities to profile unknown threats, empowering security teams to proactively disrupt anonymous adversaries targeting the airport environment \cite{kaewunruen2021digital}. ML can be applied to clusters of anomalies to generate threat profiles, providing insights into their intent, capabilities, and potential origins. These profiles are continuously refined based on the emergence of new artifacts and investigation insights, ensuring they remain dynamic and reflective of adversaries' evolving tactics \cite{naka2024optimising}.\\ By comparing profiles to known cybercriminal groups, nation-states, and insider threats, analytics and investigation focus can be prioritized effectively. Time-series and network flow analysis enable the attribution of anomalous behavior to specific internal or external sources. Through the distillation of TTPs from hunt findings and their fusion with open source intelligence, threat profiles are strengthened. Language analysis, semantic patterns, and metadata in stolen or leaked data provide valuable insights into target selection criteria and objectives. The application of AI to search unstructured dark web data further enriches the profiles, identifying mentions of airport operations, weaknesses, or plans. Predictive analytics assist in determining adversaries' technical sophistication, operational security habits, and likely next moves, enabling preemptive actions. Continuous sharing of profiles with the intelligence community fosters joint discovery of known and new threat actors. These advanced profiling capabilities arm security teams with deep insights, enabling them to proactively disrupt and mitigate the activities of anonymous adversaries targeting the airport environment.

\section{Impact Prediction and Recovery Coordination}
An airport DT plays a crucial role in enabling impact prediction \footnote{Impact prediction involves identifying the scale or severity of possible effects to provide a basis for evaluating their importance. This process aims to determine the magnitude of potential consequences.} and recovery coordination by employing proactive measures and data-driven insights \cite{luo4806351high}. It continuously monitors all vital systems and utilizes simulations to predict the potential impacts of various disruptions before they occur. When incidents happen, the DT runs dynamic simulations based on the nature and scope of the event, analyzing cascading effects and dependencies to forecast operational, financial, and reputational consequences \cite{camunez2021digital}.\\ These impact assessments 
are shared with decision-makers to inform priorities for containment and recovery efforts, while also identifying trigger points that require escalated coordination based on data-driven insights. As the recovery process unfolds, the twin models alternative approaches, tracks progress against goals, and detects emerging strains, facilitating coordinated multi-agency responses. By serving as a centralized dashboard, the DT fosters alignment across teams, reduces finger-pointing, and optimizes resource allocation in real-time. Furthermore, the twin conducts detailed simulations of the actual response to capture lessons learned, contributing to the continual refinement of resilient incident management processes \footnote{Incident management is the process employed by IT operations and DevOps teams to handle unforeseen events that may impact service quality or operations \cite{keskin2022architecting}. Its goal is to promptly identify and resolve issues while ensuring normal service and minimizing business impact.}. This evidence-based approach enables proactive and data-driven recovery management, minimizing airport disruption and downtime while ensuring effective and efficient response and recovery operations.

\subsection{Assessing Potential Attack Consequences}
A DT can play a vital role in assessing the potential consequences of cyberattacks by utilizing advanced modeling and analysis capabilities. By modeling dependency trees, the twin can foresee how system compromises could cascade through interconnected technological and business operations, providing insights into the extent of the impact. It can also simulate various attack scenarios, such as ransomware, data leaks, or infrastructure sabotage, to forecast the differing levels of functionality that may be affected.\\ The DT can project the financial costs resulting from lost ticket sales, cargo delays, rescheduling, compensation claims, and restoration efforts, enabling a comprehensive assessment of the potential financial impact. Furthermore, it can estimate the reputational fallout factors, including passenger trust, airline partnerships, and investor confidence, based on the type and scale of the incident. Workforce implications, such as absenteeism, decreased productivity, or replacement costs, can be predicted if Human Resource (HR) systems are impacted. The twin can also assess the risks to safety, national security, or environmental compliance in the event of disruptions to industrial control or emergency response systems. Additionally, it can forecast the secondary effects on the regional economy, considering the loss of jobs, tourism, and local business revenue that may result from airport operations being compromised. By utilizing attack intelligence and historical data, the DT can refine its predictions based on the most likely adversary TTPs and objectives. It continuously updates its models to adapt to new vulnerabilities and evolving threat landscapes, providing an informed basis for tailored risk management strategies and resilient architecture investments. This comprehensive approach enables organizations to make informed decisions and allocate resources effectively to mitigate the potential consequences of cyberattacks.

\subsection{Optimizing Recovery Protocols and Resources}
A DT offers valuable capabilities to optimize recovery protocols and resource allocation after an incident, enabling agile and data-driven recovery operations that are aligned across stakeholders \cite{luo4806351high,piras2024digital}. By simulating different recovery options and timelines based on damage assessments, the twin can identify the fastest pathways to restoration \cite{ma2021digital}. It models resource needs over time, allowing for optimal allocation of staff, equipment, and support according to evolving priorities. The twin predicts pinch points and tensions that could impede progress if left unaddressed, enabling proactive mitigation through coordinated efforts. Temporary workaround strategies, such as reduced schedules or cargo prioritization, can be tested if full functionality cannot be immediately restored \cite{brucherseifer2021digital}.\\ The DT facilitates the distribution of updated response protocols, incorporating lessons learned from the actual event response, through a coordinated communication network. It tracks Key Performance Indicators (KPIs) such as open issues and mean-time to repair against established goals, enabling focused efforts on areas that require attention and the removal of roadblocks \cite{petrov2023digital,kurscheidtsmart}. The twin can recommend phased restoration plans that minimize additional disruption risks arising from interim system states \cite{brusa2020digital}. It also forecasts emerging constraints from the supply chain and vendors, allowing for the activation of contingencies and contingency vendors as needed. Furthermore, the DT continuously optimizes resource allocation in real-time based on dynamic impact simulations. Through these capabilities, the DT facilitates agile and data-driven recovery operations, ensuring effective coordination and optimized resource utilization among all stakeholders involved.

\subsection{Coordinated Response Validation Across Agencies}
An airport DT can greatly facilitate coordinated response validation across multiple agencies by leveraging its capabilities for modeling, simulation, and collaboration. The twin can model the interdependencies between airport operations and external response organizations, such as emergency services, police, and the Federal Aviation Administration (FAA), to understand the intricate relationships and interactions \cite{siddiqui2024digital}. \\It can simulate handoffs, workflows, and information flows across these agencies to ensure smooth coordination during an incident. Through tabletop simulations involving all stakeholders, coordination procedures can be tested, and any gaps in roles and responsibilities can be identified \cite{yiu2021digital}. The DT can validate external agency notification protocols and escalation criteria by integrating alerts within the shared twin. It can also enable joint decision-making exercises, where virtual scenarios require multilateral approvals and resource allocation. The readiness of external partners to fulfill support functions can be assessed by activating their assets within the simulated environment \cite{luo4806351high}. The twin can further verify that external response Standard Operating Procedures (SOPs) \cite{schmitt2023business,wong2023smart,tran2022applications}.\footnote{SOP is a set of instructions created by an organization to guide workers in performing routine tasks. It ensures efficiency, quality, and consistency while minimizing errors and non-compliance with regulations.} align with airport plans through synchronized digital collaboration. After an incident, full multi-agency responses can be replayed in the DT, allowing all partners to evaluate their cooperation and conduct thorough after-action reviews. The feedback collected from these exercises can be used to continually refine integrated plans and strengthen the coordination capabilities of all agencies involved. During live responses, the airport DT can provide a unified common operating picture through a centralized information dashboard, ensuring that all stakeholders have access to real-time updates and insights. Through these collaborative efforts, the DT helps confirm that plans achieve desired outcomes by fostering aligned multi-fold stakeholder collaboration and coordination.

\section{Airport Digital Twins and eXplainable AI (XAI)}
The outcome of complex systems, such as DTs in airports and factories, is significantly influenced by human behaviors, errors, awareness levels, and response \cite{fan2021disaster,taj2022towards,lv2022artificial}. Cybersecurity, in particular, fundamentally depends on coordinated human-technology solutions. Combining DTs with XAI allows for the realization of significant benefits. Twins generate vast amounts of data, which can be analyzed by ML models. XAI complements this process by providing explanations for the outputs and recommendations of these models \cite{kobayashi2024explainable}. Trustworthiness \footnote{ An attributes that gives others trust and assurance in someone's qualifications, abilities, and dependability to carry out specific tasks and fulfill assigned responsibilities.} is a crucial characteristic of XAI, empowering the functionality of DTs \cite{suhail2023enigma,carlevarodigital}.\\
Within a DT simulation, XAI plays a vital role in providing explanations when AI models detect anomalies or potential failures. These explanations help operators understand the root causes behind issues, enabling them to take appropriate actions and effectively address real-world problems \cite{suhail2023enigma}. Additionally, XAI enhances the transparency and trustworthiness of AI predictions by highlighting the factors and scenarios that contribute to specific outcomes \cite{sarker2024explainable}.
XAI becomes invaluable when AI optimizes processes or suggests design improvements within the DT. It verifies that the models behave as expected and do not introduce unintended biases. This ensures that the proposed optimizations align with the desired goals and objectives. During the development of the DT and its associated AI models, XAI facilitates the validation of simulation inputs, such as historical data and component specifications, ensuring their proper representation and consideration within the models. This validation process enhances the reliability and accuracy of the DT's predictions and recommendations.
Explanations provided by XAI also play a crucial role in confidently transferring optimized policies or processes from simulations to real-world applications within the DT environment.\\
The transparency and interpretability offered by XAI enable operators to have a clear understanding of the optimized strategies and facilitate their implementation in practical scenarios. Furthermore, XAI enhances the debugability, auditability, and oversight of complex AI and DT systems used for mission-critical applications, such as manufacturing and infrastructure. It provides insights into the decision-making processes of the AI models, allowing for better identification and resolution of issues, as well as improved monitoring and 
control of the overall system.\\
Real-world implementations of XAI in DT airport systems have potential for showing significant improvements in security, operational efficiency, maintenance practices, passenger experience, and air traffic management \cite{kong2024explainable}. For security and threat detection, XAI is employed to analyze surveillance data and passenger behavior within the DT airport system, providing explanations for AI model alerts or flagged events. This transparency empowers security operators to make informed decisions and take effective actions to address security concerns.\\
In terms of operational efficiency, XAI optimizes airport operations within the DT environment by providing explanations for the AI model's recommendations on process improvements, resource allocation, and scheduling adjustments. This enables operators to understand the factors and scenarios influencing the AI-driven optimizations, resulting in fine-tuned operations, enhanced efficiency, and data-driven decision-making.
Predictive maintenance is another area where XAI proves valuable in DT airport systems. AI models analyze sensor data from critical airport assets to predict maintenance needs and potential failures, with XAI techniques providing explanations for the AI model's predictions. Maintenance teams can then understand the root causes and proactively plan maintenance activities, reducing downtime and optimizing asset performance.
Moreover, XAI enhances the passenger experience within DT airport systems by analyzing passenger data and providing personalized recommendations and assistance. XAI techniques explain the AI model's suggestions, allowing passengers to understand the reasons behind specific recommendations or notifications. This transparency builds trust and confidence in the AI-driven enhancements, leading to improved customer satisfaction.\\
In the context of air traffic management, XAI is applied to optimize various aspects such as routing, congestion management, and scheduling strategies \cite{degas2022survey}. AI models analyze air traffic data, weather conditions, and airport capacity, with XAI techniques providing explanations for the AI model's recommendations. This transparency empowers air traffic controllers to make informed choices and ensure safe and efficient air traffic operations.
Overall, XAI plays a crucial role in enhancing security measures, operational efficiency, maintenance practices, passenger experience, and air traffic management within DT airport systems. By providing interpretable explanations, XAI empowers operators, security personnel, maintenance teams, air traffic controllers, and passengers to leverage the insights, trustworthiness and recommendations generated by AI models, leading to improved airport operations and services.
\subsection{Unveiling the Depth of Explainable AI (XAI) in Digital Twins}
XAI is a flourishing field with numerous common algorithms and approaches \cite{minh2022explainable}. This subsection bridges the gap between these common XAI approaches and their potential applications in the context of DTS. Some of the well known XAI algorithms and approaches are Local Interpretable Model-agnostic Explanations (LIME)/Shapley Values, Attention Mechanisms (AM), Anchors, Concept Activation Vectors (CAVs), Layer-wise Relevance Propagation (LRP), Interpretable Embeddings, Decision Trees, Model Distillation, Counterfactual Explanations, Influence Functions and Causal Inference.\\
A taxonomy of XAI techniques for DT can categorize them based on explanation type, explanation form, model aspect explained, model agnosticism, fidelity, granularity, and purpose as can be seen in the Figure . 
LIME/Shapley Values explain individual predictions by approximating them locally with an interpretable model like linear regression. Attention Mechanisms visualize which parts of the input text/image were "attended" to for predictions in Transformers and Convolutional Neural Network (CNN) respectively. Anchors identify minimal sets of conditions that are individually necessary for a prediction; CAVs visualize which concepts/semantic meanings were associated with predictions in text classifiers.\\ LRP traces back prediction relevance to input features for image and language models. Interpretable Embeddings learn low-dimensional, semantically meaningful embeddings to explain text/image encodings. Decision Trees are self-explanatory rule-based models that are inherently interpretable. Model Distillation trains a transparent, interpretable "student" model to mimic a complex "teacher" model. Counterfactual Explanations suggest the smallest plausible changes to input to flip the prediction. Influence Functions quantify the effect of individual training examples on model parameters and predictions. Causal Inference explains predictive relationships using causal graphs and reasoning about interventions. The choice of technique depends on the type of model, data modality, and level of interpretability required, and combining complementary techniques also improves explanations.\\
Each of the above approaches and algorithms can be applied to DTs. Attention mechanisms can help visualize factors like weather and flight delays that influenced departure time predictions. Anchors can aid in understanding conditions like passenger volumes and gate assignments required for on-time arrival forecasts. CAVs can interpret textual factors like safety incidents and regulations considered by simulators. LRP can trace back runway/taxiway condition predictions in aerial imagery models to specific visual features; Interpretable embeddings can help in understanding passenger profiles learned from itineraries and demographics for personalization.\\ Decision trees can self-explain screening rules and maintenance schedules from historical data without black boxes. Model distillation can simplify complex multi-agent crowd simulations for staff training and communication. Counterfactuals can determine the sensitivity of gate planning to small changes in flight schedules and demands. Influence functions can assess the effect of past incidents like drone sightings on predictive systems. Causal graphs can represent relationships between factors like weather and passenger volumes for "what-if" analysis, providing transparency, justification, and troubleshooting capabilities for critical airport DT systems.

\section{Algorithms used for Enhancing Airport Cybersecurity with Digital Twin Solutions}
DTs and algorithms are pivotal in bolstering cybersecurity measures at airports \cite{susila2020impact}. A DT serves as a virtual replica of an airport's IT systems, networks, and infrastructure, facilitating simulations to analyze potential cyber threats without endangering live operational systems. This allows for testing various attack vectors, payloads, and malware to understand their impacts and develop appropriate responses.\\
Algorithms play a crucial role within the DT environment, autonomously detecting threats, monitoring network traffic patterns, analyzing user behaviors, and identifying anomalies that indicate security breaches. ML algorithms profile normal network traffic, while behavioral analysis algorithms monitor user and device behaviors. Network flow analysis algorithms examine traffic patterns \cite{faleiro2022digital}.
While the DT operates virtually, it can integrate with actual airport systems to some extent, enabling hybrid testing where the twin interacts with real systems while containing impacts within the virtual model.\\
Insights gained from DT analysis help identify cybersecurity weaknesses, prioritize enhancements, develop contingency plans, and train incident response teams. Continuously feeding real-world data into the DT enhances the intelligence of models and algorithms over time, improving threat detection and response capabilities.
The DT enables the development and testing of a range of algorithms to enhance airport security \cite{li2022novel}. Anomaly detection algorithms compare real-time network traffic to baseline models, detect unexpected changes, and identify suspicious patterns. Behavioral analysis algorithms, such as User Entity Behavior Analysis (UEBA) and Device Entity Behavior Analysis (DEBA), profile behaviors to detect breaches \cite{xu2021digital}. Anomaly detection algorithms, such as GANs, learn normal patterns and detect anomalies or suspicious behavior in the digital twin for security purposes. \\
Threat intelligence algorithms automate vulnerability scanning and check against the latest threat feeds \cite{nintsiou2023threat}. Threat hunting algorithms empower security teams to proactively search for indicators of compromise or unusual behaviors \cite{karaarslan2021digital}. Domain name and IP reputation checking algorithms provide real-time checks against blacklists.
For malware detection, algorithms analyze files for known signatures, inspect running processes for suspicious behaviors, and filter traffic for malware patterns \cite{cha2023intelligent}.
Network security algorithms optimize rulesets for firewalls and Intrusion Prevention Systems (IPS) \cite{wang2024cyber}. Software-Defined Networking (SDN) security configurations experiment with dynamic network micro-segmentation. Zero trust network access algorithms model fine-grained access policies.
By utilizing a DT, airports can develop, train, optimize, and evaluate advanced algorithms in a safe environment, strengthening cyber defenses and safeguarding critical systems against evolving threats.\\
In addition, GenAI algorithms can be applied in an airport DT. Generative modeling using algorithms like GANs helps generate novel and optimized concepts for facilities, layouts, and traffic flows \cite{vakaruk2021digital}. Procedural content generation algorithms automatically generate high-fidelity 3D assets for the DT, such as buildings and vehicles \cite{xu2021digital,hasan2023wasserstein}. Sim2Real training uses generative models to synthesize simulated twin data for pre-training other AI systems, overcoming data limitations \cite{li2022sim2real}. Generative forecasting algorithms provide accurate time-series predictions, including passenger volumes, delays, and maintenance needs based on historical patterns. Conditional generation enables the incorporation of specific constraints or what-if scenarios, such as generating evacuation plans under certain safety rules. Optimization via evolution employs genetic algorithms and other generative techniques to coevolve and optimize complex airport systems \cite{hadar2020cyber}. 

\section{Future Applications and Technologies}
In the era of advanced GenAI capabilities like deepfakes, DTs are set to play a crucial role in fortifying airports' cyber defenses and ensuring the safety of operations and passengers.
The future applications of airport DTs are likely to incorporate the use of generative models to simulate increasingly realistic cyber threats \cite{singh2021digital}. This would enable security teams to test their capabilities in detecting and responding to novel and sophisticated attacks, enhancing their readiness \cite{empl2024digital}.\\
Moreover, DTs can leverage generative AI to autonomously generate synthetic airport data and systems, creating a controlled test environment to identify vulnerabilities and uncover new attack vectors before they pose a risk to real operations \cite{boje2020towards}. This proactive approach allows for the development of robust defense strategies and the implementation of effective countermeasures.\\
The integration of advanced AI threat modeling within DTs is instrumental in predicting the evolution of cyber risks associated with technologies like deepfakes. By staying ahead of emerging threats, airports can continuously improve their detection methods, forensic analysis techniques, and overall resilience strategies \cite{bhatti2021towards}.\\
GenAI can also enhance DTs' capabilities in anomaly detection by synthesizing normal behaviors and outliers. This augmentation enables them to identify sophisticated intrusions and insider threats more effectively, providing an additional layer of protection against cyberattacks.
As the attack surface expands and cyber threats become more sophisticated, DTs will continue to serve as centralized cyber planning and coordination hubs \cite{wang2022mobility}. Their role in safeguarding airport operations and ensuring passenger safety is poised to grow in importance throughout the generative AI era and beyond.

\subsection{Integrating Generative Design for Enhanced Security}
Integrating generative design capabilities into airport DTs offers numerous ways to enhance security and stay ahead of evolving threats in the future. By utilizing generative models, the twins can explore vast design spaces and recommend optimized system architectures that minimize known vulnerabilities, ensuring robust security measures are in place \cite{jimmy2024pioneering}. They can automatically generate diverse synthetic variants of network configurations, assets, and data to test the resistance against new and emerging threats, allowing for proactive defense strategies \cite{muhammad2024integrating}. DTs can also develop generative defense blueprints that outline multi-layered protections tailored to the capabilities of suspected adversaries, providing a comprehensive security framework \cite{wang2023survey}. Moreover, they can prototype deliberately deceptive environments such as honeypots, canaries, and moving targets to misdirect adversaries or detect intrusions, adding an additional layer of security \cite{nintsiou2023threat}. The twins can generate synthetically derived red team tools and new attack methodologies, continuously strengthening defenses by anticipating future risks and vulnerabilities. \\ Additionally, DTs can dynamically reshape airport infrastructure within virtual simulations to evaluate design trade-offs and future-proof security measures preemptively \cite{kor2023investigation}. They can continuously regenerate security controls like monitoring rules, access policies, and encryption in response to emerging best practices or regulatory requirements. By generating just-in-time access credentials and environment configurations, the twins enhance internal segmentation and reduce long-term vulnerabilities. Finally, autonomously remediating discovered issues becomes possible by designing and testing patches or workarounds within the DT before deploying them in the real-world environment. Through these generatively-enhanced DT defenses, airports can proactively adapt to ongoing technological changes and innovation in threat landscapes, ensuring robust and resilient security measures are in place.

\subsection{Modeling Emergent Technologies Like IoT/OT}
Airport DTs have the potential to model and integrate emerging technologies like IoT/OT in the future, enabling smooth and secure implementation within complex airport infrastructure \cite{padovano2024improving}. By generating digital representations of new IoT/OT devices and systems, the twins can capture their characteristics and functionalities accurately. They can continuously simulate interactions between legacy and emerging technologies to identify potential integration issues or new attack surfaces, ensuring seamless compatibility. The DTs can model the cascading impacts of vulnerabilities in IoT/OT systems, allowing for the forecasting of secondary consequences across interconnected operations and enabling proactive risk mitigation \cite{haleem2024perspective}. \\
Through the twins, layers of protective instrumentation, monitoring, and responsive controls for emerging technologies can be thoroughly tested before live deployment, ensuring robust security measures are in place. Synthetic network traffic patterns involving IoT/OT can be generated to test analytics for detecting compromised devices and uncovering potential threats. The DTs can autonomously patch and reconfigure simulated IoT/OT environments, evaluating mitigations and ensuring that security is incorporated into digital designs from the very beginning. They can anticipate the long-term evolution of IoT/OT technologies and the associated risk profiles over their lifecycles, facilitating progressive planning and adaptation to changing scenarios \cite{santos2024ai}. The DTs also serve as a platform for multi-disciplinary collaboration between technology, security, and airport teams, enabling them to collectively address the human and technical challenges associated with IoT/OT integration. DTs can continuously expand to represent emerging innovations at the conceptual stage, allowing for the proactive inclusion of safety and resilience considerations in their designs. Through these capabilities, airport DTs can play a crucial role in smoothly and securely integrating disruptive technologies into complex airport infrastructure, ensuring a seamless and resilient operational environment.

\subsection{Expanding Collaborative Defense Through Twins}
In the future, airport DTs have the potential to expand collaborative defense capabilities by leveraging globally distributed yet interconnected networks of twins \cite{singh2021digital}. These DTs can facilitate real-time coordination between multiple interconnected airport twins, enabling the rapid sharing of indicators, detecting distributed threats, and synchronizing responses across different regions \cite{becue2020new}. By integrating with national-level cyberthreat sharing platforms, the DTs can strengthen public-private intelligence cooperation, enhancing the collective ability to combat persistent and advanced adversaries \cite{alcaraz2022digital}.\\
Additionally, partnering with research communities allows the twins to leverage crowd-sourced generative red teaming, anonymous simulation testing of mitigations, and contribute to advancements in the field of digital infrastructure defense. To foster seamless collaboration, standardized twin ecosystems and APIs can be developed, enabling airports, vendors, and agencies to exchange analytic models, anomalies, and playbooks, collectively enhancing defenses \cite{fuller2020digital}. The DTs can also coordinate response collaboration at scale through federated learning across decentralized twin networks, autonomously fortifying individual entities and strengthening systemic resilience. Furthermore, establishing dynamic Cyber Mutual Aid (CMA) frameworks enables twin-enabled entities to automatically request and provide surge assistance during acute incidents that surpass their individual response capacities \cite{lampropoulos2023enhancing}. To enhance international coordination, virtual simulations of collaborative incidents and exercises can be conducted, strengthening synchrony of cross-border coordination for threats that disregard physical and legal jurisdictions. By multiplying the combined cyber defense potential through globally distributed yet coherently linked digital twin networks, airports can significantly enhance their defense capabilities and ensure a robust and resilient security posture.\\

\subsection{Advantages of Agentic AI for Airport Digital Twins}
One of the next generations of GenAI is called Agentic AI \cite{zhukov2024agentic}.
Agentic AI describes AI systems that have the ability to act independently to accomplish predetermined objectives \cite{shavit2023practices}. In contrast to conventional AI which demands continuous human supervision, agentic AI can make decisions, adjust to new data, and evolve by itself based on real-world experiences over time \cite{khanliquidity}. Rather than needing constant human oversight, these goal-driven AI agents are endowed with autonomy to navigates real complex environments and complete designated tasks through self-directed actions and learning.\\
Agentic AI may have the benefit over generative AI for an airport DT in its ability to provide autonomy and adaptability.
Agentic AI can independently respond and adapt to changes in airport conditions, such as weather delays or equipment outages, in real-time without human intervention. This ability to autonomously problem-solve and distribute decision-making is crucial for effectively operating a highly dynamic system like an airport.
Another advantage is the goal-directed behavior that agentic AI enables. By configuring the DT with agentic AI agents, each pursuing designated objectives (such as maximizing on-time departures, minimizing fuel usage), complex interactions across different airport functions can be coordinated towards strategic goals. Generative models alone may not exhibit the same level of purposeful and optimized behavior.\\
Agentic AI trained through Reinforcement Learning also offers improved simulation accuracy. It can incrementally enhance its actions based on realistic digital consequences, incorporating feedback that better simulates cause-and-effect in the physical airport system over many iterative simulations. This leads to more reliable predictions compared to generative models.\\
Furthermore, agentic AI demonstrates adaptability to new situations. Since it is goal-driven rather than solely reliant on pattern-matching, it excels in generalizing to new scenarios not present in its training data. This allows it to effectively respond to unexpected events, ensuring the digital twin remains functional even without internet connectivity to retrain central models.
Lastly, agentic AI is well-suited for interoperability in simulating a complex system with independent components, such as different terminal operations and security checkpoints. These components must interact and coordinate decisions in real-time based on local information. Distributed agentic AI facilitates this coordination, enabling efficient and realistic simulations within the airport digital twin.

\section{Conclusions}
This paper explored how airport DT powered by advanced GenAI and ML including explainable AI (XAI), techniques can significantly enhance airport cybersecurity preparedness and defenses. DTs enable comprehensive vulnerability assessment through virtual modeling of realistic cyberattack scenarios and generation of synthetic threat data.
They facilitate prioritized remediation through predictive risk analysis and accelerated validation of fixes. DTs also support hands-on staff training with realistic simulation environments. Tabletop exercises allow planning of coordinated incident response protocols. Impact prediction and recovery coordination functions optimize resource allocation and validation of multi-agency response plans.
Anomaly detection and threat hunting are strengthened by leveraging AI to proactively monitor networks and identify intrusion behaviors. Future applications outlined show even greater security benefits, such as integrating generative design, modeling emerging technologies like IoT/OT, and collaborative defense through interconnected twins. Training algorithms with XAI will optimize human integration with DT platforms.\\
As threats to airports increase in scale and sophistication, DT solutions powered by advanced AI, including XAI, demonstrate tremendous promise for enhancing cybersecurity. DTs provide unified situational awareness, predictive analytics, and coordinated inter-agency capabilities to address critical preparedness and resilience gaps. Their continuous monitoring, anomaly detection, impact prediction and collaboration ensure effective detection and response to evolving cyber threats.
Embracing DT technology is crucial for airports to anticipate and mitigate risks from disruptive innovations. It will help safely integrate emerging technologies while staying ahead of the changing cyber landscape. DTs establish a new frontier in achieving the highest levels of airport cybersecurity through synergistic human-AI relationships enabled by this transformative technology. Overall, DTs show significant potential to revolutionize airport cyber defenses and ensure safety amid evolving threats.

\bibliographystyle{IEEEtran}
\bibliography{ref.bib}

\begin{thebibliography}{100}
\providecommand{\url}[1]{#1}
\csname url@samestyle\endcsname
\providecommand{\newblock}{\relax}
\providecommand{\bibinfo}[2]{#2}
\providecommand{\BIBentrySTDinterwordspacing}{\spaceskip=0pt\relax}
\providecommand{\BIBentryALTinterwordstretchfactor}{4}
\providecommand{\BIBentryALTinterwordspacing}{\spaceskip=\fontdimen2\font plus
\BIBentryALTinterwordstretchfactor\fontdimen3\font minus \fontdimen4\font\relax}
\providecommand{\BIBforeignlanguage}[2]{{%
\expandafter\ifx\csname l@#1\endcsname\relax
\typeout{** WARNING: IEEEtran.bst: No hyphenation pattern has been}%
\typeout{** loaded for the language `#1'. Using the pattern for}%
\typeout{** the default language instead.}%
\else
\language=\csname l@#1\endcsname
\fi
#2}}
\providecommand{\BIBdecl}{\relax}
\BIBdecl

\bibitem{lykou2018smart}
G.~Lykou, A.~Anagnostopoulou, and D.~Gritzalis, ``Smart airport cybersecurity: Threat mitigation and cyber resilience controls,'' \emph{Sensors}, vol.~19, no.~1, p.~19, 2018.

\bibitem{kuleshov2024cyber}
Y.~A. Kuleshov, K.~Nagpal, K.~Ucpinar, A.~Gadaginmath, S.~Gadaginmath, K.~O’Daniel, D.~Sun, L.~Tan, N.~Veatch, and H.~Monangi, ``Cyber attacks on avionics networks in digital twin environment: Detection and defense,'' in \emph{AIAA SCITECH 2024 Forum}, 2024, p. 0277.

\bibitem{grieves2017digital}
M.~Grieves and J.~Vickers, ``Digital twin: Mitigating unpredictable, undesirable emergent behavior in complex systems,'' \emph{Transdisciplinary perspectives on complex systems: New findings and approaches}, pp. 85--113, 2017.

\bibitem{xu2024leveraging}
H.~Xu, F.~Omitaomu, S.~Sabri, X.~Li, and Y.~Song, ``Leveraging generative ai for smart city digital twins: A survey on the autonomous generation of data, scenarios, 3d city models, and urban designs,'' \emph{arXiv preprint arXiv:2405.19464}, 2024.

\bibitem{vanderhorn2021digital}
E.~VanDerHorn and S.~Mahadevan, ``Digital twin: Generalization, characterization and implementation,'' \emph{Decision support systems}, vol. 145, p. 113524, 2021.

\bibitem{lyu2024handbook}
Z.~Lyu, \emph{Handbook of Digital Twins}.\hskip 1em plus 0.5em minus 0.4em\relax CRC Press, 2024.

\bibitem{feuerriegel2024generative}
S.~Feuerriegel, J.~Hartmann, C.~Janiesch, and P.~Zschech, ``Generative ai,'' \emph{Business \& Information Systems Engineering}, vol.~66, no.~1, pp. 111--126, 2024.

\bibitem{airportir}
H.~Dorries, ``Will airports' digital twins be the next big thing?'' \url{https://airportir.com/ir-pulse/will-airports-digital-twins-be-the-next-big-thing}, 2023, [Online; accessed 30-July-2024].

\bibitem{elmarady2022impact}
A.~A. ElMarady and K.~H. Rahouma, ``The impact of covid-19 on the cybersecurity in civil aviation: Review and analysis,'' in \emph{2022 International Telecommunications Conference (ITC-Egypt)}.\hskip 1em plus 0.5em minus 0.4em\relax IEEE, 2022, pp. 1--6.

\bibitem{koroniotis2020holistic}
N.~Koroniotis, N.~Moustafa, F.~Schiliro, P.~Gauravaram, and H.~Janicke, ``A holistic review of cybersecurity and reliability perspectives in smart airports,'' \emph{IEEE Access}, vol.~8, pp. 209\,802--209\,834, 2020.

\bibitem{auer2019cyber}
M.~E. Auer \emph{et~al.}, \emph{Cyber-physical Systems and Digital Twins: Proceedings of the 16th International Conference on Remote Engineering and Virtual Instrumentation}.\hskip 1em plus 0.5em minus 0.4em\relax Springer, 2019, vol.~80.

\bibitem{senarak2024port}
C.~Senarak, ``Port cyberattacks from 2011 to 2023: a literature review and discussion of selected cases,'' \emph{Maritime Economics \& Logistics}, vol.~26, no.~1, pp. 105--130, 2024.

\bibitem{zhiling2024construction}
H.~Zhiling, Z.~Luyao, C.~Qian, T.~Xin, and L.~Xiaohuan, ``Construction and analysis of a taxiing conflict prediction model in airport scene based on digital twin,'' \emph{Journal of Xihua University (Natural Science Edition)}, vol.~43, no.~1, pp. 8--15, 2024.

\bibitem{nath2021building}
S.~V. Nath, P.~Van~Schalkwyk, and D.~Isaacs, \emph{Building industrial digital twins: Design, develop, and deploy digital twin solutions for real-world industries using Azure digital twins}.\hskip 1em plus 0.5em minus 0.4em\relax Packt Publishing Ltd, 2021.

\bibitem{digitalplatform}
D.~T. Consortium \emph{et~al.}, ``Platform stack architectural framework: An introductory guide form, 2023.''

\bibitem{mcclenaghan2023cost}
A.~McClenaghan, J.~Gopsill, R.~Ballantyne, and B.~Hicks, ``Cost benefit analysis for digital twin model selection at the time of investment.'' \emph{Procedia CIRP}, vol. 120, pp. 1197--1202, 2023.

\bibitem{yin2023state}
Y.~Yin, P.~Zheng, C.~Li, and L.~Wang, ``A state-of-the-art survey on augmented reality-assisted digital twin for futuristic human-centric industry transformation,'' \emph{Robotics and Computer-Integrated Manufacturing}, vol.~81, p. 102515, 2023.

\bibitem{lv2021artificial}
Z.~Lv and S.~Xie, ``Artificial intelligence in the digital twins: State of the art, challenges, and future research topics [version 1; peer review,'' 2021.

\bibitem{uzun2019design}
M.~Uzun, M.~U. Demirezen, E.~Koyuncu, and G.~Inalhan, ``Design of a hybrid digital-twin flight performance model through machine learning,'' in \emph{2019 IEEE Aerospace conference}.\hskip 1em plus 0.5em minus 0.4em\relax IEEE, 2019, pp. 1--14.

\bibitem{saifutdinov2020digital}
F.~Saifutdinov, I.~Jackson, J.~Tolujevs, and T.~Zmanovska, ``Digital twin as a decision support tool for airport traffic control,'' in \emph{2020 61st International Scientific Conference on Information Technology and Management Science of Riga Technical University (ITMS)}.\hskip 1em plus 0.5em minus 0.4em\relax IEEE, 2020, pp. 1--5.

\bibitem{cheng2023review}
R.~Cheng, L.~Hou, and S.~Xu, ``A review of digital twin applications in civil and infrastructure emergency management,'' \emph{Buildings}, vol.~13, no.~5, p. 1143, 2023.

\bibitem{ding2024generative}
W.~Ding, ``Generative ai for critical digital twins,'' Ph.D. dissertation, Carnegie Mellon University, 2024.

\bibitem{souanef2023digital}
T.~Souanef, S.~Al-Rubaye, A.~Tsourdos, S.~Ayo, and D.~Panagiotakopoulos, ``Digital twin development for the airspace of the future,'' \emph{Drones}, vol.~7, no.~7, p. 484, 2023.

\bibitem{su2023using}
H.~Su and W.~Pan, ``Using digital twins to integrate cyber security with physical security at smart airports,'' \emph{Interdisciplinary Journal of Engineering and Environmental Sciences}, vol.~10, no.~1, pp. 38--45, 2023.

\bibitem{koroniotis2021sair}
N.~Koroniotis, N.~Moustafa, F.~Schiliro, P.~Gauravaram, and H.~Janicke, ``The sair-iiot cyber testbed as a service: A novel cybertwins architecture in iiot-based smart airports,'' \emph{IEEE Transactions on Intelligent Transportation Systems}, vol.~24, no.~2, pp. 2368--2381, 2021.

\bibitem{papadopoulos2024protection}
L.~Papadopoulos, K.~Demestichas, E.~Mu{\~n}oz-Navarro, J.~J. Hern{\'a}ndez-Montesinos, S.~Paul, N.~Museux, S.~K{\"o}nig, S.~Schauer, A.~C. Alarc{\'o}n, I.~P. Llopis \emph{et~al.}, ``Protection of critical infrastructures from advanced combined cyber and physical threats: The praetorian approach,'' \emph{International Journal of Critical Infrastructure Protection}, vol.~44, p. 100657, 2024.

\bibitem{xiong2022digital}
M.~Xiong and H.~Wang, ``Digital twin applications in aviation industry: A review,'' \emph{The International Journal of Advanced Manufacturing Technology}, vol. 121, no.~9, pp. 5677--5692, 2022.

\bibitem{stoddart2022cyberwar}
K.~Stoddart, ``Cyberwar: Attacking critical infrastructure,'' in \emph{Cyberwarfare: Threats to Critical Infrastructure}.\hskip 1em plus 0.5em minus 0.4em\relax Springer, 2022, pp. 147--225.

\bibitem{yiu2021digital}
C.~Y. Yiu, K.~K. Ng, C.-H. Lee, C.~T. Chow, T.~C. Chan, K.~C. Li, and K.~Y. Wong, ``A digital twin-based platform towards intelligent automation with virtual counterparts of flight and air traffic control operations,'' \emph{Applied Sciences}, vol.~11, no.~22, p. 10923, 2021.

\bibitem{conde2022applying}
J.~Conde, A.~Munoz-Arcentales, M.~Romero, J.~Rojo, J.~Salvach{\'u}a, G.~Huecas, and {\'A}.~Alonso, ``Applying digital twins for the management of information in turnaround event operations in commercial airports,'' \emph{Advanced Engineering Informatics}, vol.~54, p. 101723, 2022.

\bibitem{brucherseifer2021digital}
E.~Brucherseifer, H.~Winter, A.~Mentges, M.~M{\"u}hlh{\"a}user, and M.~Hellmann, ``Digital twin conceptual framework for improving critical infrastructure resilience,'' \emph{at-Automatisierungstechnik}, vol.~69, no.~12, pp. 1062--1080, 2021.

\bibitem{piroumian2023cybersecurity}
V.~Piroumian, ``Cybersecurity and dependability for digital twins and the internet of things,'' in \emph{The Digital Twin}.\hskip 1em plus 0.5em minus 0.4em\relax Springer, 2023, pp. 365--393.

\bibitem{li2021digital}
L.~Li, S.~Aslam, A.~Wileman, and S.~Perinpanayagam, ``Digital twin in aerospace industry: A gentle introduction,'' \emph{IEEE Access}, vol.~10, pp. 9543--9562, 2021.

\bibitem{faleiro2022digital}
R.~Faleiro, L.~Pan, S.~R. Pokhrel, and R.~Doss, ``Digital twin for cybersecurity: Towards enhancing cyber resilience,'' in \emph{Broadband Communications, Networks, and Systems: 12th EAI International Conference, BROADNETS 2021, Virtual Event, October 28--29, 2021, Proceedings 12}.\hskip 1em plus 0.5em minus 0.4em\relax Springer, 2022, pp. 57--76.

\bibitem{wen2022toward}
J.~Wen, B.~Gabrys, and K.~Musial, ``Toward digital twin oriented modeling of complex networked systems and their dynamics: A comprehensive survey,'' \emph{Ieee Access}, vol.~10, pp. 66\,886--66\,923, 2022.

\bibitem{kaewunruen2021digital}
S.~Kaewunruen, J.~Sresakoolchai, W.~Ma, and O.~Phil-Ebosie, ``Digital twin aided vulnerability assessment and risk-based maintenance planning of bridge infrastructures exposed to extreme conditions,'' \emph{Sustainability}, vol.~13, no.~4, p. 2051, 2021.

\bibitem{padovano2024improving}
A.~Padovano, F.~Longo, L.~Manca, and R.~Grugni, ``Improving safety management in railway stations through a simulation-based digital twin approach,'' \emph{Computers \& Industrial Engineering}, vol. 187, p. 109839, 2024.

\bibitem{borras2020d1}
M.~Borr{\`a}s and J.~L. IDP, ``D1. 6--ethics \& privacy, information security,'' 2020.

\bibitem{gao2022adequate}
J.~Gao, W.~Wu \emph{et~al.}, ``Adequate testing unmanned autonomous vehicle systems-infrastructures, approaches, issues, challenges, and needs,'' in \emph{2022 IEEE International Conference on Service-Oriented System Engineering (SOSE)}.\hskip 1em plus 0.5em minus 0.4em\relax IEEE, 2022, pp. 154--164.

\bibitem{sadeghi2024digital}
A.~Sadeghi, P.~Bellavista, W.~Song, and M.~Yazdani-Asrami, ``Digital twins for condition and fleet monitoring of aircraft: Towards more-intelligent electrified aviation systems,'' \emph{IEEE Access}, 2024.

\bibitem{nsoh2021next}
J.~Nsoh, ``“next-gen” cybersecurity,'' 2021.

\bibitem{hazbon2019digital}
O.~Hazbon, L.~Gutierrez, C.~Bil, M.~Napolitano, and M.~Fravolini, ``Digital twin concept for aircraft system failure detection and correction,'' in \emph{AIAA Aviation 2019 Forum}, 2019, p. 2887.

\bibitem{kilic2023digital}
U.~Kilic, G.~Yalin, and O.~Cam, ``Digital twin for electronic centralized aircraft monitoring by machine learning algorithms,'' \emph{Energy}, vol. 283, p. 129118, 2023.

\bibitem{reitenbach2020collaborative}
S.~Reitenbach, M.~Vieweg, R.~Becker, C.~Hollmann, F.~Wolters, J.~Schmeink, T.~Otten, and M.~Siggel, ``Collaborative aircraft engine preliminary design using a virtual engine platform, part a: Architecture and methodology,'' in \emph{AIAA Scitech 2020 Forum}, 2020, p. 0867.

\bibitem{ruiz2024application}
A.~Ruiz de~la Torre~Acha, R.~M. Rio~Belver, J.~Fernandez~Aguirrebe{\~n}a, and C.~Merlo, ``Application of simulation and virtual reality to production learning,'' \emph{Education+ Training}, vol.~66, no. 2/3, pp. 145--165, 2024.

\bibitem{wong2023smart}
E.~T. Wong and W.~Man, ``Smart maintenance and human factor modeling for aircraft safety,'' in \emph{Applications in Reliability and Statistical Computing}.\hskip 1em plus 0.5em minus 0.4em\relax Springer, 2023, pp. 25--59.

\bibitem{hristov2023challenges}
G.~Hristov, I.~Beloev, and P.~Zahariev, ``Challenges, requirements, opportunities and solutions for the digital transformation of the transport education.'' \emph{Strategies for Policy in Science \& Education/Strategii na Obrazovatelnata i Nauchnata Politika}, vol.~31, 2023.

\bibitem{tran2022applications}
T.~H. Tran, Y.~Jiang, and L.~Williams, ``Applications of mixed reality for smart aviation industry: Opportunities and challenges,'' \emph{Modern Development and Challenges in Virtual Reality}, 2022.

\bibitem{wong2021closed}
E.~Y. Wong, D.~Y. Mo, and S.~So, ``Closed-loop digital twin system for air cargo load planning operations,'' \emph{International Journal of Computer Integrated Manufacturing}, vol.~34, no. 7-8, pp. 801--813, 2021.

\bibitem{ojo2024innovative}
B.~Ojo, J.~C. Ogborigbo, and M.~O. Okafor, ``Innovative solutions for critical infrastructure resilience against cyber-physical attacks,'' 2024.

\bibitem{applebystrategic}
J.~Appleby, A.~Bitoun, H.~ten Bergen, F.~Legras, and A.-G. Bosser, ``The strategic project: Innovative automated analysis tools for after action review (aar) using ai and modeling \& simulation.''

\bibitem{luo4806351high}
M.~Luo, H.~Fricke, B.~Desart, S.~R. Zapata, and M.~Schultz, ``High-fidelity digital twin applied agent-based model for supporting predictable airport ground operations,'' \emph{Available at SSRN 4806351}.

\bibitem{meyer2020development}
H.~Meyer, J.~Zimdahl, A.~Kamtsiuris, R.~Meissner, F.~Raddatz, S.~Haufe, and M.~B{\"a}{\ss}ler, ``Development of a digital twin for aviation research,'' 2020.

\bibitem{agapaki2022airport}
E.~Agapaki, ``Airport digital twins for resilient disaster management response,'' in \emph{International Conference on Learning and Intelligent Optimization}.\hskip 1em plus 0.5em minus 0.4em\relax Springer, 2022, pp. 467--486.

\bibitem{ariyachandra2023digital}
M.~M.~F. Ariyachandra and G.~Wedawatta, ``Digital twin smart cities for disaster risk management: a review of evolving concepts,'' \emph{Sustainability}, vol.~15, no.~15, p. 11910, 2023.

\bibitem{casey2024real}
L.~Casey, J.~Dooley, M.~Codd, R.~Dahyot, M.~Cognetti, T.~Mullarkey, P.~Redmond, and G.~Lacey, ``A real-time digital twin for active safety in an aircraft hangar,'' \emph{Frontiers in Virtual Reality}, vol.~5, p. 1372923, 2024.

\bibitem{sindiramutty2023autonomous}
S.~R. Sindiramutty, ``Autonomous threat hunting: A future paradigm for ai-driven threat intelligence,'' \emph{arXiv preprint arXiv:2401.00286}, 2023.

\bibitem{wang2023digital}
W.~Wang, Q.~Zaheer, S.~Qiu, W.~Wang, C.~Ai, J.~Wang, S.~Wang, and W.~Hu, ``Digital twins technologies,'' in \emph{Digital Twin Technologies in Transportation Infrastructure Management}.\hskip 1em plus 0.5em minus 0.4em\relax Springer, 2023, pp. 27--74.

\bibitem{hakiri2024comprehensive}
A.~Hakiri, A.~Gokhale, S.~B. Yahia, and N.~Mellouli, ``A comprehensive survey on digital twin for future networks and emerging internet of things industry,'' \emph{Computer Networks}, p. 110350, 2024.

\bibitem{khan2022digital}
L.~U. Khan, Z.~Han, W.~Saad, E.~Hossain, M.~Guizani, and C.~S. Hong, ``Digital twin of wireless systems: Overview, taxonomy, challenges, and opportunities,'' \emph{IEEE Communications Surveys \& Tutorials}, vol.~24, no.~4, pp. 2230--2254, 2022.

\bibitem{susila2020impact}
N.~Susila, A.~Sruthi, and S.~Usha, ``Impact of cloud security in digital twin,'' in \emph{Advances in Computers}.\hskip 1em plus 0.5em minus 0.4em\relax Elsevier, 2020, vol. 117, no.~1, pp. 247--263.

\bibitem{naka2024optimising}
K.~K.~L. Naka, ``Optimising airport security: Biometric sorting by departure time,'' Ph.D. dissertation, Swinburne University, 2024.

\bibitem{camunez2021digital}
A.~Cam{\'u}{\~n}ez~i Guirao, ``Digital transformation at airports: the impact of the bim and the iot technologies on the airport environment,'' Master's thesis, Universitat Polit{\`e}cnica de Catalunya, 2021.

\bibitem{keskin2022architecting}
B.~Keskin, B.~Salman, and O.~Koseoglu, ``Architecting a bim-based digital twin platform for airport asset management: a model-based system engineering with sysml approach,'' \emph{Journal of Construction Engineering and Management}, vol. 148, no.~5, p. 04022020, 2022.

\bibitem{piras2024digital}
G.~Piras, S.~Agostinelli, and F.~Muzi, ``Digital twin framework for built environment: a review of key enablers,'' \emph{Energies}, vol.~17, no.~2, p. 436, 2024.

\bibitem{ma2021digital}
F.~Ma, ``Digital twin at urban scale for a master plan in the area of the cebu airport (philippine),'' Ph.D. dissertation, Politecnico di Torino, 2021.

\bibitem{petrov2023digital}
S.~Petrov, ``Digital twins and sustainability: A comprehensive review of limitations and opportunities,'' 2023.

\bibitem{kurscheidtsmart}
R.~J. Kurscheidt~Netto, E.~d.~F. Rocha~Loures, and E.~A. Portela~Santos, ``Smart scheduled industrial maintenance: A joint scheduling maintenance and production framework using multicriteria optimization based on process mining and delay-time modeling,'' \emph{Eduardo Alves, Smart Scheduled Industrial Maintenance: A Joint Scheduling Maintenance and Production Framework Using Multicriteria Optimization Based on Process Mining and Delay-Time Modeling}.

\bibitem{brusa2020digital}
E.~Brusa, ``Digital twin: Toward the integration between system design and rams assessment through the model-based systems engineering,'' \emph{IEEE Systems Journal}, vol.~15, no.~3, pp. 3549--3560, 2020.

\bibitem{siddiqui2024digital}
F.~M. Siddiqui and C.~J. Mead, ``Digital twin conops as a platform for airport master planning,'' in \emph{2024 New Trends in Civil Aviation (NTCA)}.\hskip 1em plus 0.5em minus 0.4em\relax IEEE, 2024, pp. 105--111.

\bibitem{schmitt2023business}
L.~Schmitt and D.~Copps, ``The business of digital twins,'' in \emph{The Digital Twin}.\hskip 1em plus 0.5em minus 0.4em\relax Springer, 2023, pp. 21--63.

\bibitem{fan2021disaster}
C.~Fan, C.~Zhang, A.~Yahja, and A.~Mostafavi, ``Disaster city digital twin: A vision for integrating artificial and human intelligence for disaster management,'' \emph{International journal of information management}, vol.~56, p. 102049, 2021.

\bibitem{taj2022towards}
I.~Taj and N.~Zaman, ``Towards industrial revolution 5.0 and explainable artificial intelligence: Challenges and opportunities,'' \emph{International Journal of Computing and Digital Systems}, vol.~12, no.~1, pp. 295--320, 2022.

\bibitem{lv2022artificial}
Z.~Lv and S.~Xie, ``Artificial intelligence in the digital twins: State of the art, challenges, and future research topics,'' \emph{Digital Twin}, vol.~1, p.~12, 2022.

\bibitem{kobayashi2024explainable}
K.~Kobayashi and S.~B. Alam, ``Explainable, interpretable, and trustworthy ai for an intelligent digital twin: A case study on remaining useful life,'' \emph{Engineering Applications of Artificial Intelligence}, vol. 129, p. 107620, 2024.

\bibitem{suhail2023enigma}
S.~Suhail, M.~Iqbal, R.~Hussain, and R.~Jurdak, ``Enigma: An explainable digital twin security solution for cyber--physical systems,'' \emph{Computers in Industry}, vol. 151, p. 103961, 2023.

\bibitem{carlevarodigital}
A.~Carlevaro, G.~De~Bernardi, M.~Lenatti, S.~Narteni, M.~Muselli, A.~Paglialonga, F.~Dabbene, and M.~Mongelli, ``Are digital twins suitable to drive safe ai?''

\bibitem{sarker2024explainable}
I.~H. Sarker, H.~Janicke, A.~Mohsin, A.~Gill, and L.~Maglaras, ``Explainable ai for cybersecurity automation, intelligence and trustworthiness in digital twin: Methods, taxonomy, challenges and prospects,'' \emph{ICT Express}, 2024.

\bibitem{kong2024explainable}
X.~Kong, Y.~Xing, A.~Tsourdos, Z.~Wang, W.~Guo, A.~Perrusquia, and A.~Wikander, ``Explainable interface for human-autonomy teaming: A survey,'' \emph{arXiv preprint arXiv:2405.02583}, 2024.

\bibitem{degas2022survey}
A.~Degas, M.~R. Islam, C.~Hurter, S.~Barua, H.~Rahman, M.~Poudel, D.~Ruscio, M.~U. Ahmed, S.~Begum, M.~A. Rahman \emph{et~al.}, ``A survey on artificial intelligence (ai) and explainable ai in air traffic management: Current trends and development with future research trajectory,'' \emph{Applied Sciences}, vol.~12, no.~3, p. 1295, 2022.

\bibitem{minh2022explainable}
D.~Minh, H.~X. Wang, Y.~F. Li, and T.~N. Nguyen, ``Explainable artificial intelligence: a comprehensive review,'' \emph{Artificial Intelligence Review}, pp. 1--66, 2022.

\bibitem{li2022novel}
Z.~Li, M.~Duan, B.~Xiao, and S.~Yang, ``A novel anomaly detection method for digital twin data using deconvolution operation with attention mechanism,'' \emph{IEEE Transactions on Industrial Informatics}, vol.~19, no.~5, pp. 7278--7286, 2022.

\bibitem{xu2021digital}
Q.~Xu, S.~Ali, and T.~Yue, ``Digital twin-based anomaly detection in cyber-physical systems,'' in \emph{2021 14th IEEE Conference on Software Testing, Verification and Validation (ICST)}.\hskip 1em plus 0.5em minus 0.4em\relax IEEE, 2021, pp. 205--216.

\bibitem{nintsiou2023threat}
M.~Nintsiou, E.~Grigoriou, P.~A. Karypidis, T.~Saoulidis, E.~Fountoukidis, and P.~Sarigiannidis, ``Threat intelligence using digital twin honeypots in cybersecurity,'' in \emph{2023 IEEE International Conference on Cyber Security and Resilience (CSR)}.\hskip 1em plus 0.5em minus 0.4em\relax IEEE, 2023, pp. 530--537.

\bibitem{karaarslan2021digital}
E.~Karaarslan and M.~Babiker, ``Digital twin security threats and countermeasures: An introduction,'' in \emph{2021 International Conference on Information Security and Cryptology (ISCTURKEY)}.\hskip 1em plus 0.5em minus 0.4em\relax IEEE, 2021, pp. 7--11.

\bibitem{cha2023intelligent}
H.-J. Cha, H.-K. Yang, Y.-J. Song, and A.~R. Kang, ``Intelligent anomaly detection system through malware image augmentation in iiot environment based on digital twin,'' \emph{Applied Sciences}, vol.~13, no.~18, p. 10196, 2023.

\bibitem{wang2024cyber}
Z.~Wang, ``Cyber digital twin with deep learning model for enterprise products management,'' \emph{Wireless Personal Communications}, pp. 1--30, 2024.

\bibitem{vakaruk2021digital}
S.~Vakaruk, A.~Mozo, A.~Pastor, and D.~R. L{\'o}pez, ``A digital twin network for security training in 5g industrial environments,'' in \emph{2021 IEEE 1st International Conference on Digital Twins and Parallel Intelligence (DTPI)}.\hskip 1em plus 0.5em minus 0.4em\relax IEEE, 2021, pp. 395--398.

\bibitem{hasan2023wasserstein}
M.~N. Hasan, S.~U. Jan, and I.~Koo, ``Wasserstein gan-based digital twin-inspired model for early drift fault detection in wireless sensor networks,'' \emph{IEEE Sensors Journal}, vol.~23, no.~12, pp. 13\,327--13\,339, 2023.

\bibitem{li2022sim2real}
P.~Li, J.~Thomas, X.~Wang, H.~Erdol, A.~Ahmad, R.~Inacio, S.~Kapoor, A.~Parekh, A.~Doufexi, A.~Shojaeifard \emph{et~al.}, ``Sim2real for reinforcement learning driven next generation networks,'' \emph{arXiv preprint arXiv:2206.03846}, 2022.

\bibitem{hadar2020cyber}
E.~Hadar, D.~Kravchenko, and A.~Basovskiy, ``Cyber digital twin simulator for automatic gathering and prioritization of security controls’ requirements,'' in \emph{2020 IEEE 28th International Requirements Engineering Conference (RE)}.\hskip 1em plus 0.5em minus 0.4em\relax IEEE, 2020, pp. 250--259.

\bibitem{singh2021digital}
M.~Singh, E.~Fuenmayor, E.~P. Hinchy, Y.~Qiao, N.~Murray, and D.~Devine, ``Digital twin: Origin to future,'' \emph{Applied System Innovation}, vol.~4, no.~2, p.~36, 2021.

\bibitem{empl2024digital}
P.~EMPL, D.~KOCH, M.~DIETZ, and G.~PERNUL, ``Digital twins in security operations: State of the art and future perspectives,'' 2024.

\bibitem{boje2020towards}
C.~Boje, A.~Guerriero, S.~Kubicki, and Y.~Rezgui, ``Towards a semantic construction digital twin: Directions for future research,'' \emph{Automation in construction}, vol. 114, p. 103179, 2020.

\bibitem{bhatti2021towards}
G.~Bhatti, H.~Mohan, and R.~R. Singh, ``Towards the future of smart electric vehicles: Digital twin technology,'' \emph{Renewable and Sustainable Energy Reviews}, vol. 141, p. 110801, 2021.

\bibitem{wang2022mobility}
Z.~Wang, R.~Gupta, K.~Han, H.~Wang, A.~Ganlath, N.~Ammar, and P.~Tiwari, ``Mobility digital twin: Concept, architecture, case study, and future challenges,'' \emph{IEEE Internet of Things Journal}, vol.~9, no.~18, pp. 17\,452--17\,467, 2022.

\bibitem{jimmy2024pioneering}
S.~Jimmy and K.~Berhane, ``Pioneering the future: Generative ai and digital twin integration in 6g networks,'' \emph{Authorea Preprints}, 2024.

\bibitem{muhammad2024integrating}
K.~Muhammad, T.~David, G.~Nassisid, and T.~Farus, ``Integrating generative ai with network digital twins for enhanced network operations,'' \emph{arXiv preprint arXiv:2406.17112}, 2024.

\bibitem{wang2023survey}
Y.~Wang, Z.~Su, S.~Guo, M.~Dai, T.~H. Luan, and Y.~Liu, ``A survey on digital twins: Architecture, enabling technologies, security and privacy, and future prospects,'' \emph{IEEE Internet of Things Journal}, vol.~10, no.~17, pp. 14\,965--14\,987, 2023.

\bibitem{kor2023investigation}
M.~Kor, I.~Yitmen, and S.~Alizadehsalehi, ``An investigation for integration of deep learning and digital twins towards construction 4.0,'' \emph{Smart and Sustainable Built Environment}, vol.~12, no.~3, pp. 461--487, 2023.

\bibitem{haleem2024perspective}
A.~Haleem, M.~Javaid, and R.~P. Singh, ``Perspective of leadership 4.0 in the era of fourth industrial revolution: A comprehensive view,'' \emph{Journal of Industrial Safety}, p. 100006, 2024.

\bibitem{santos2024ai}
O.~Santos, S.~Salam, and H.~Dahir, ``The ai revolution in networking, cybersecurity, and emerging technologies,'' 2024.

\bibitem{becue2020new}
A.~B{\'e}cue, E.~Maia, L.~Feeken, P.~Borchers, and I.~Pra{\c{c}}a, ``A new concept of digital twin supporting optimization and resilience of factories of the future,'' \emph{Applied Sciences}, vol.~10, no.~13, p. 4482, 2020.

\bibitem{alcaraz2022digital}
C.~Alcaraz and J.~Lopez, ``Digital twin: A comprehensive survey of security threats,'' \emph{IEEE Communications Surveys \& Tutorials}, vol.~24, no.~3, pp. 1475--1503, 2022.

\bibitem{fuller2020digital}
A.~Fuller, Z.~Fan, C.~Day, and C.~Barlow, ``Digital twin: Enabling technologies, challenges and open research,'' \emph{IEEE access}, vol.~8, pp. 108\,952--108\,971, 2020.

\bibitem{lampropoulos2023enhancing}
G.~Lampropoulos and K.~Siakas, ``Enhancing and securing cyber-physical systems and industry 4.0 through digital twins: A critical review,'' \emph{Journal of software: evolution and process}, vol.~35, no.~7, p. e2494, 2023.

\bibitem{zhukov2024agentic}
V.~Zhukov, ``What is agentic ai? understanding agentic ai,'' \url{https://zhukov.live/what-is-agentic-ai-understanding-agentic-ai-5f011521bc08}, 2024, [Online; accessed 30-July-2024].

\bibitem{shavit2023practices}
Y.~Shavit, S.~Agarwal, M.~Brundage, S.~Adler, C.~O’Keefe, R.~Campbell, T.~Lee, P.~Mishkin, T.~Eloundou, A.~Hickey \emph{et~al.}, ``Practices for governing agentic ai systems,'' \emph{Research Paper, OpenAI, December}, 2023.

\bibitem{khanliquidity}
I.~U. Khan, A.~Khan, and A.~Matyana, ``Liquidity is all you need: A decentralized approach to generative agentic ai systems with liquidity.''

\end{thebibliography}

\end{document}